\documentclass[fleqn,twoside]{article}
\usepackage{espcrc2}

\usepackage{graphicx}
\usepackage{epsfig}
\usepackage[figuresright]{rotating}
\usepackage{bm}
\usepackage{amssymb}
\newcommand{\AmS}{{\protect\the\textfont2
  A\kern-.1667em\lower.5ex\hbox{M}\kern-.125emS}}

\title{$\alpha_s$ From Hadronic $\tau$ Decay Data}

\author{K. Maltman
\address{Dept. of Mathematics and Statistics, York Univ.,
4700 Keele St., Toronto, ON CANADA M3J 1P3}%
        \thanks{Alternate address: CSSM, University of Adelaide, Adelaide
SA 5005 Australia},
     T. Yavin\address{Dept. of Physics and Astronomy, York Univ.,
4700 Keele St., Toronto, ON CANADA M3J 1P3}}
       
\begin{document}

\begin{abstract}
We discuss the extraction of $\alpha_s$ using isovector hadronic 
$\tau$ decay data and sum rules constructed specifically to suppress 
contributions associated with poorly known higher dimension condensates.
We show, first, that problems with the treatment of such contributions affect
earlier related analyses and, second, that these problems can be brought under
good theoretical control through the use of an alternate analysis strategy.
Our results, run up to the $n_f=5$ regime, correspond to
$\alpha_s(M_Z^2)=0.1187\pm 0.0016$, in excellent agreement with
the recently updated global fit to electroweak data at the Z scale 
and other high-scale direct determinations.
\vspace{1pc}
\end{abstract}

% typeset front matter (including abstract)
\maketitle

\section{Introduction and Background}
The strong coupling, $\alpha_s$, at some conventionally chosen
reference scale, is one of the fundamental parameters of the 
Standard Model (SM). Its value, in the $n_f=3$ regime, can be extracted
using hadronic $\tau$ decay data as a consequence of the
finite energy sum rule (FESR) relation 
\begin{equation}
\int_0^{s_0}w(s)\, \rho(s)\, ds\, =\, -{\frac{1}{2\pi i}}\oint_{\vert
s\vert =s_0}w(s)\, \Pi (s)\, ds
\label{basicfesr}
\end{equation}
which is valid for any analytic weight, $w(s)$, and any correlator $\Pi (s)$ 
without kinematic singularities. In Eq.~(\ref{basicfesr}), $\rho (s)$ is the 
spectral function of $\Pi (s)$. The basic idea is to use experimental spectral 
data on the LHS and, for sufficiently large $s_0$, the OPE representation 
of $\Pi$ (which involves $\alpha_s$) on the RHS. The region of 
applicability of the OPE is extended to lower $s_0$ when $w(s)$ satisfies
the condition $w(s=s_0)=0$, which suppresses contributions on the RHS from the 
region of the contour near the timelike real axis~\cite{pqw,kmfesr}. 

Experimental input for the LHS of Eq.~(\ref{basicfesr})
is available because, in the SM,
the kinematics of $\tau$ decay allows the inclusive rate
for decays mediated by the flavor $ij=ud,us$, vector (V) or 
axial vector (A) hadronic currents to be written as
kinematically weighted integrals over the spectral functions 
$\rho_{V/A;ij}^{(J)}(s)$ of the spin $J=0,1$ components 
of the relevant current-current two-point functions~\cite{tsaitaubasic}. 
Explicitly, with $y_\tau\, \equiv\, s/m_\tau^2$ and
$R_{V/A;ij}\equiv {\frac{\Gamma [\tau^- \rightarrow \nu_\tau
\, {\rm hadrons}_{V/A;ij}\, (\gamma)]}{\Gamma [\tau^- \rightarrow
\nu_\tau e^- {\bar \nu}_e (\gamma)]}}$, we have
%~\cite{additiveewfootnote}
\begin{eqnarray}
&&R_{V/A;ij}= 12\pi^2\vert V_{ij}\vert^2 S_{EW}\,
\int^{1}_0\, dy_\tau \, \left( 1-y_\tau\right)^2 \nonumber\\
&&\ \left[ \left( 1 + 2y_\tau\right)
\rho_{V/A;ij}^{(0+1)}(s) - 2y_\tau \rho_{V/A;ij}^{(0)}(s) \right]
\label{taukinspectral}
\end{eqnarray}
with $V_{ij}$ the flavor $ij$ CKM matrix element, $S_{EW}$ a short-distance 
electroweak (EW) correction, and $\rho_{V/A;ij}^{(0+1)}(s)
\equiv\rho_{V/A;ij}^{(1)}(s)+\rho_{V/A;ij}^{(0)}(s)$. 
%We will restrict our attention to $ij=ud$ in what follows.

For $ij=ud$, apart from the $\pi$ contribution to $\rho_{A;ud}^{(0)}$, 
$\rho_{V;ud}^{(0)}(s)$ and $\rho_{A;ud}^{(0)}(s)$ are numerically negligible, 
being proportional to $O([m_d\mp m_u]^2)$. The sum of flavor $ud$ V and 
A spectral functions, $\rho_{V+A;ud}^{(0+1)}(s)$, can thus
be extracted from the differential decay distribution $dR_{V+A;ud}/ds$,
for all $s<m_\tau^2\simeq 3.16\ {\rm GeV}^2$. 
Further separation into V and A components is unambiguous 
for $n\pi$ states, but requires additional input for 
$K\bar{K}n\pi$ ($n>0$) states, making errors on the experimental
distribution smallest for the V+A sum. 

Given the spectral functions $\rho^{0+1}_{V,A,V+A;ud}$, FESRs for the related 
correlators, $\Pi^{(0+1)}_{T;ud}$, with $T=V,A,V+A$, are straightforwardly 
constructable. For the scales $s_0\gtrsim 2\ {\rm GeV}^2$ 
considered here, the OPE representations of these correlators
are strongly dominated by the dimension $D=0$ contribution, which is entirely
determined by $\alpha_s$, converges well, and is known to 
$O(\alpha_s^4)$~\cite{bck08}. The resulting FESRs are thus well
adapted to the determination of $\alpha_s$. To optimize the precision of 
this determination, however, care must be taken in evaluating the
small, residual higher $D$ contributions, a $\sim 1\%$ determination of 
$\alpha_s(M_Z^2)$, for example, requiring control of higher $D$ contributions 
at the level of $\sim 0.5\%$ of the $D=0$ term~\cite{my08}.

For $ij=ud$ and $s_0\gtrsim 2\ {\rm GeV}^2$, $D=2$ contributions are 
numerically negligible, being either $O(m_{u,d}^2)$ or 
$O(\alpha_s^2 m_s^2)$~\cite{chkw93}. $D=4$ contributions are,
to very good accuracy, determined by the RG invariant condensates
$\langle m_\ell \bar{\ell}\ell\rangle$,
$\langle m_s \bar{s}s\rangle$ and $\langle aG^2\rangle$,
for which phenomenological input exists
(see Ref.~\cite{bnp} for the explicit forms of these contributions,
and Ref.~\cite{my08} for details of the condensate values employed).
$D\geq 6$ contributions are more problematic since the relevant condensates 
are poorly known or phenomenologically undetermined. We deal with these
contributions by defining effective condensate combinations, $C_6,C_8,\cdots$, 
such that
$\left[\Pi (Q^2)\right]^{OPE}_{D>4}\, \equiv\, \sum_{D=6,8,\cdots}
C_D/Q^D$
(up to logarithmic corrections) and fitting these quantities
to data. This process is greatly facilitated by working with
polynomial weights $w(s)=\sum_{m=0} b_m y^m$ defined in terms of
the dimensionless variable $y=s/s_0$. For such weights, the integrated
$D\geq 6$ OPE contributions have the form 
\begin{eqnarray}
&&{\frac{-1}{2\pi i}}\oint_{\vert s\vert =s_0}
ds\, w(y)\,
\left[\Pi (Q^2)\right]^{OPE}_{D>4}\nonumber\\
&&\qquad\qquad =\, 
\sum_{k=2}(-1)^kb_k{\frac{C_{2k+2}}{s_0^k}}
\label{higherd}\end{eqnarray}
allowing contributions of different $D$ to be distinguished by their
differing $s_0$ dependences.

\section{Problems With Existing Analyses}
Existing analyses are based on the approach pioneered by 
ALEPH and OPAL~\cite{alephud9798,opalud99}. In this approach, OPE contributions
with $D>8$ are assumed safely negligible for all weights employed,
and the quantities $\alpha_s(m_\tau^2)$, $\langle aG^2\rangle$,
$C_6$ and $C_8$ are fitted using the $s_0=m_\tau^2$ values of the
spectral integrals corresponding to the 
$(km)=(00),(10),(11),(12)$ and $(13)$ ``$(km)$ spectral weights'',
$w^{(km)}(y)=(1-y)^ky^m w^{(00)}(y)$, where $w^{(00)}(y)=(1-y)^2(1+2y)$ 
is the kinematic weight occuring on the RHS of Eq.~(\ref{taukinspectral}). 
ALEPH~\cite{alephud9798,finalalpehdhzreview,davieretal08} performed 
this fit independently for each of the V, A and V+A channels, while
OPAL~\cite{opalud99} performed independent fits for the V+A and 
combined V,A channels. A potential problem with the assumption
that all $D>8$ contributions can be safely neglected is the
fact that $w^{(km)}$ has degree $3+k+m$ which, from Eq.~(\ref{higherd}),
implies that contributions with $D$ up to $16$ are, in principle,
present in at least one of the FESRs considered in the 
ALEPH and OPAL analyses. 
Since only the single $s_0$ value $s_0=m_\tau^2$ is
employed, there is no way to prevent the fit from adjusting to
the presence of any neglected, but non-negligible, $D>8$ contributions 
by shifting the lower $D$ parameters determined in the fit in
such a way as to compensate, as best as possible, for the missing terms.
Such a problem with the fit can only be exposed by studying
the same, or related, FESRs over a range of $s_0$, where the
different scaling with $s_0$ of terms of different $D$
will become operative.

A simple way to test for the presence (or absence) of such problems in the
ALEPH and OPAL fits is to consider the $s_0$-dependent fit-qualities,
\begin{equation}
F^w_T(s_0)\equiv {\frac{I^w_{spec}(s_0)
-I^w_{OPE}(s_0)}{\delta I_{spec}^w(s_0)}}
\label{fitqdefn}\end{equation}
where $T=V, A$ or $V+A$, $I_{OPE}^w(s_0)$ and $I_{spec}^w(s_0)$ 
are the OPE and spectral integrals appearing, respectively, on the RHS 
and LHS of the corresponding $w(s)$-weighted FESR, 
and $\delta I^w_{spec}(s_0)$ is the error on $I_{spec}^w(s_0)$, determined 
using the experimental covariance matrix for $dR_{T;ud}/ds$. Because
of strong correlations between values corresponding to the same $w(s)$
but different $s_0$, a fitted version of the OPE representation 
should be considered
reliable only if $\vert F^w_T(s_0)\vert$ remains $\lesssim 1$
for a range of $s_0<m_\tau^2$. 

\begin{figure*}
\unitlength1cm
\caption{Comparison of the $F_V^w(s_0)$ corresponding to (i) our fits
and (ii) the 2005 ALEPH fit, and various weights having degree
$\leq 3$. The light (heavy) 
dotted line corresponds to the ALEPH fit (our fit) for $w^{(00)}$,
the light (heavy) dashed line to the ALEPH fit (our fit) for 
$w_2$, the light (heavy) dot-dashed line to the ALEPH fit (our fit)
for $w_3$, and the light (heavy) double-dot-dashed line to
the ALEPH fit (our fit) for $w(y)=y(1-y)^2$.}
\rotatebox{270}{\mbox{
\begin{minipage}[thb]{11.7cm}
\begin{picture}(11.6,15.1)
\epsfig{figure=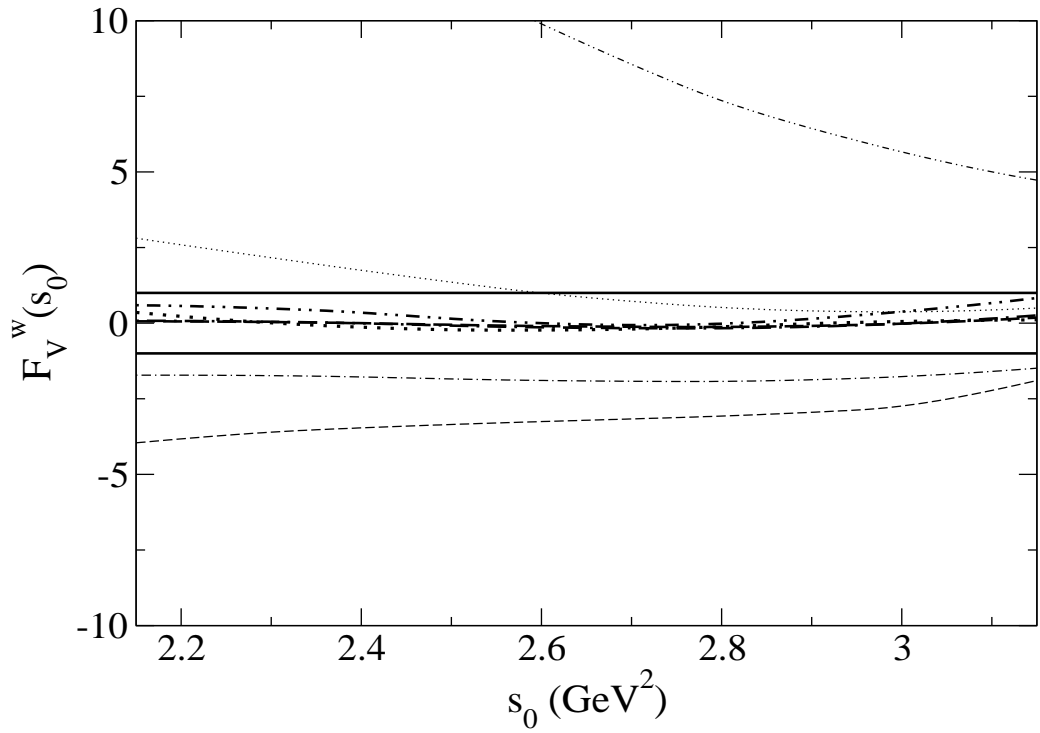,height=15.5cm,width=11.5cm}
\end{picture}
\end{minipage}}}
\label{fig1}\end{figure*}

In Ref.~\cite{my08} this condition was shown to be far from satisfied for any
of the spectral weights employed in the ALEPH and OPAL analyses. An 
illustration of the problem is provided in Figure~\ref{fig1}, which shows
the $F_V^w(s_0)$ corresponding to the 2005 ALEPH final data and
OPE fit for four weights, $w^{(00)}(y)$,
$w_2(y)=(1-y)^2$, $w_3(y)=1-{\frac{3}{2}}y+{\frac{y^3}{2}}$, and
$w(y)=y(1-y)^2$, all having degree $\leq 3$ (and hence 
OPE contributions only up to $D=8$). If the fitted values
for the $D\leq 8$ OPE parameters obtained by ALEPH are reliable, 
one should find an $s_0$ window below $m_\tau^2$ having
$\vert F^w_V(s_0)\vert \lesssim 1$ for all four weights.
The results, given by the light lines in the figure,
show that no such window exists. In fact, for the weights $w_2$,
$w_3$ and $y(1-y)^2$ not employed in the original fit, the fit
quality is poor even at $s_0=m_\tau^2$.

The problem seen in the Figure could be due either to contamination
of the $D\leq 8$ OPE fit parameters by neglected, but non-negligible,
$D>8$ contributions, or to OPE breakdown. One may test the latter possibility 
by performing alternate fits in which potential $D>8$
contributions, where present, are explicitly taken into account.
The results of such fits, discussed in the next section, yield alternate 
OPE representations in excellent agreement with the corresponding spectral 
integrals over a range of $s_0$, both for the weights employed
in the fits and for related weights with OPE representations
determined by the same set of OPE parameters. The resulting
alternate fit qualities, $F_V^w(s_0)$, for the four degree $\leq 3$ weights
already discussed above, are shown by the dark lines in Figure~\ref{fig1}.
The results clearly show no evidence for OPE breakdown.
%We now turn to the details underlying these alternate fit results.

\section{Alternate FESR Analyses}
In what follows, we employ the updated charmonium sum rule 
determination of $\langle aG^2\rangle$~\cite{newgcond4}. 
The gluon condensate term dominates the $D=4$ OPE contribution. 
Details on the input for the 
small corrections proportional to $\langle m_\ell \bar{\ell}\ell\rangle$
and $\langle m_s \bar{s}s\rangle$ may be found in Ref.~\cite{my08}.

The $D=0$ contributions are evaluated using the expression for the
$D=0$ Adler function series from Ref.~\cite{bck08}.
An $O(\alpha_s^5)$ contribution, employing the estimated value 
for the corresponding coefficient from Ref.~\cite{bck08}, 
is included for our central fit.
We consider both the contour improved (CIPT) and
fixed order (FOPT) determinations of the integrated $D=0$ sum.
The reference scale $n_f=3$ coupling needed in the evaluation of the CIPT and
FOPT sums (taken to be $\alpha_s(m_\tau^2)$) is a parameter to be 
determined in the fit.

Since, for V and A correlators, 
OPE breakdown is expected (and observed) to set in for $s_0$ 
below $\sim 2\ {\rm GeV}^2$~\cite{gik01,cdgmvma}, 
a limited window of $s_0$ values is available for use in fitting
$\alpha_s$ and the unknown $C_{D>4}$. It is thus convenient 
to work with FESRs based on the weights
\begin{equation}
w_N(y)\, =\, 1\, -\, {\frac{N}{N-1}}\, y\, +\, {\frac{1}{N-1}}\, y^N,
\label{wNdefn}\end{equation}
which, like $w^{(00)}(y)$, have a double zero at $s=s_0$ ($y=1$).
From Eq.~(\ref{higherd}) we see that the $w_N$ FESR involves
only a single integrated unknown $D>4$ OPE contribution, 
${\frac{(-1)^N}{(N-1)}}{\frac{C_{2N+2}}{s_0^N}}$. As $N$ is increased, 
the scaling of this contribution with $s_0$ becomes more and more rapid, aiding
in the fitting of $C_{2N+2}$. The decrease in the coefficient factor,
$1/(N-1)$, also means that the corresponding FESR is more strongly
$D=0$ dominated, a desirable situation for the determination of
$\alpha_s$. The latter effect is dominant
for sufficiently large $N$. In addition, as $N$ is increased,
$w_N(y)\rightarrow (1-y)$, whose single zero at $s=s_0$
provides less strong suppression of contributions
from the region of the timelike point on the OPE contour.
We thus restrict our attention to the $w_2,\cdots ,w_6$ FESRs.

Since $D=2$ contributions are negligible and $D=4$ contributions
are fixed by phenomenological input, the only OPE parameters
to be fit using the $w_N$ FESR are $\alpha_s(m_\tau^2)$ and $C_{2N+2}$.
The $C_{2N+2}$ will of course depend on the channel (V, A or V+A)
being considered. The values for $\alpha_s(m_\tau^2)$ 
obtained using the different $w_N$ and/or different channels should, however,
be consistent, and this consistency represents an important cross-check 
on the reliability of the analysis framework.

We have analyzed the $w_2,\cdots ,w_6$ FESRs using
the final 2005 ALEPH isovector data and covariances, in each of
the V, A and V+A channels. A similar analysis has been performed
for the V+A channel using the OPAL data and covariances. See
Ref.~\cite{my08} for further details, and a discussion of the
reasons for the analysis choices. 

We report here only on the
results for $\alpha_s(m_\tau^2)$. A full discussion of the
errors, and results for the $C_{2N+2}$, may be found 
in Ref.~\cite{my08}. Results obtained using the CIPT prescription
are presented in Table~\ref{table1}.
For each entry, the first error is experimental (computed using
the experimental covariance matrix, and including the $0.32\%$
normalization uncertainty), while the second is theoretical.
The corresponding FOPT-based results may be found in Ref.~\cite{my08}.

\begin{table}
\caption{\label{table1}Results of the $w_N$ CIPT-based FESR fits for
$\alpha_s(m_\tau^2)$ obtained
using either the ALEPH or OPAL data and covariances. 
The first error is experimental and the second theoretical.}
\vskip .1in
\begin{tabular}{|c|c|c|c|}
\hline
Data set&Channel&Weight&
$\alpha_s\left( m_\tau^2\right)$\\
\hline
ALEPH&V&$w_2$&\ \ $0.321(7)(12)$\\
&&$w_3$&\ \ $0.321(7)(12)$\\
&&$w_4$&\ \ $0.321(7)(12)$\\
&&$w_5$&\ \ $0.321(7)(12)$\\
&&$w_6$&\ \ $0.321(7)(12)$\\
\hline
&A&$w_2$&\ \ $0.319(6)(12)$\\
&&$w_3$&\ \ $0.319(6)(12)$\\
&&$w_4$&\ \ $0.319(6)(12)$\\
&&$w_5$&\ \ $0.319(6)(12)$\\
&&$w_6$&\ \ $0.319(6)(12)$\\
\hline
&V+A&$w_2$&\ \ $0.320(5)(12)$\\
&&$w_3$&\ \ $0.320(5)(12)$\\
&&$w_4$&\ \ $0.320(5)(12)$\\
&&$w_5$&\ \ $0.320(5)(12)$\\
&&$w_6$&\ \ $0.320(5)(12)$\\
\hline
\hline
OPAL&V+A&$w_2$&\ \ $0.322(7)(12)$\\
&&$w_3$&\ \ $0.322(7)(12)$\\
&&$w_4$&\ \ $0.322(7)(12)$\\
&&$w_5$&\ \ $0.322(7)(12)$\\
&&$w_6$&\ \ $0.322(8)(12)$\\
\hline
\end{tabular}
\end{table}

From the table we see excellent consistency between the 
ALEPH-based V, A and V+A results, as well as between the ALEPH
and OPAL V+A results. There is also extremely good consistency
within each channel between results obtained using the different $w_N$.
This consistency is realized only after
fitting the small, but non-negligible, $D>4$
contributions on the OPE sides of the various FESRs~\cite{my08}.
The FOPT results corresponding to different $w_N$ but the same channel 
display significantly less good consistency~\cite{my08}.
From the table, one sees that theoretical errors are a factor
of $\sim 2$ larger than the corresponding experimental errors.
The dominant contribution to the theoretical
error is that associated with the truncation of the series
for the dominant $D=0$ OPE contribution. Our truncation error is
the sum in quadrature of the shift in $\alpha_s$ produced
by including the last incorporated term in the truncated Adler
function series and the difference between the results produced
by the FOPT and CIPT evaluations. The latter
contribution is the larger of the two and, as noted
previously in the literature, shows no signs of decreasing with 
increasing truncation order.
The FOPT-CIPT difference thus dominates the current uncertainty.

Our central determination for $\alpha_s$ is based on the V+A analyses,
which has the smallest experimental errors. Averaging the
ALEPH- and OPAL-based results using the non-normalization
component of the experimental errors yields
\begin{equation}
\alpha_s(m_\tau^2)\, =\, 0.3209(46)(118)
\label{finalalphastau}\end{equation}
where the errors are experimental (including normalization)
and theoretical, respectively. 

The $n_f=5$ result, $\alpha_s(M_Z^2)$, is obtained from 
Eq.~(\ref{finalalphastau}) using the standard
self-consistent combination of $4$-loop running with $3$-loop matching
at the flavor thresholds~\cite{cks97}. With
$m_c(m_c)=1.286(13)$ GeV, $m_b(m_b)=4.164(25)$ GeV~\cite{kss07},
matching thresholds $rm_{c,b}(m_{c,b})$ with
$r$ varying between $0.7$ and $3$, and standard estimates for the
effect of the truncated running and matching,
the evolution-induced uncertainty on $\alpha_s(M_Z^2)$ is $0.0003$.
Our final result is then
\begin{equation}
\alpha_s(M_Z^2)\, =\, 0.1187(3)_{evol}(6)_{exp}(15)_{th}\, .
\label{finalalphasmz}\end{equation}
%where the first uncertainty is due to evolution, the second is experimental
%and the third theoretical. 
The difference between this value and that 
obtained in Ref.~\cite{davieretal08}, $0.1212(11)$, serves to quantify 
the impact of the $D>8$ contributions neglected in the earlier
spectral weight analyses.

The result, Eq.~(\ref{finalalphasmz}), is in excellent agreement
with recent independent determinations, including that of the
updated global EW analysis~\cite{bck08,davieretal08} and 
two recent updates~\cite{hpqcd08,mlms08} of the older HPQCD/UKQCD 
lattice determination~\cite{latticealphas}.

With theory errors dominant, and the $D=0$
truncation uncertainty dominating the theory error, further
improvement will be possible only if a better understanding
of the truncation uncertainty can be obtained. In a recent
paper, Beneke and Jamin~\cite{jb08} investigated this
issue using a model which incorporates the general structure associated
with the first two IR renormalon and leading UV renormalon singularities of
the resummed $D=0$ series. This particular version of
what could be a more general model
was used to argue in favor of the FOPT over the CIPT prescription.
The FOPT evaluation, as well as the resummed model, together
with assumed values for $C_6$ and $C_8$, were also used to 
determine $\alpha_s$.
While it was shown in Ref.~\cite{my08} that the assumed $C_6$ and $C_8$ 
values are not consistent with the $\alpha_s$ obtained using FOPT, 
the underlying approach remains extremely interesting. In fact,
one can see that the minimal version of the model employed in
Ref.~\cite{jb08} predicts CIPT approximations which deviate from the model
version of the related resummed series by amounts that are sizeable,
and a factor of $\sim 2$ larger for $w^{(00)}$ than for
$w_2$ and $w_3$~\cite{jaminprivate}. The deviations are significantly
smaller for the FOPT versions. Were the model to provide a good
representation of the true resummed series, one would thus
expect no set of $\alpha_s$, $C_6$ and $C_8$ to provide a good quality 
simultaneous CIPT fit for the $w_2$, $w_3$ and $w^{(00)}$ spectral
integrals, while a reasonable quality simultaneous fit would be 
expected to exist using the FOPT prescription. In fact, just the opposite 
occurs: the values of $\alpha_s$, $C_6$ and $C_8$ obtained from
a combined CIPT fit to the $w_2$, $w_3$ FESRs provide also an excellent
representation of the $w^{(00)}$ spectral integrals, while those
obtained from the corresponding FOPT fit provide a very poor one. 
This suggests an extended version of the Beneke-Jamin model,
perhaps taking into account more IR and/or UV renormalon singularities, 
is likely to be needed. The fact that it is possible
to reach such a conclusion, however, immediately makes clear that the sets
of variously weighted $s_0$-dependent spectral integrals provide 
significant constraints for use in constructing such generalizations. 
It is thus likely that such modelling can be improved in
future, and used to reduce the truncation uncertainty component
in the determination of $\alpha_s$.

\end{document}